\documentclass[sn-aps]{sn-jnl}
\usepackage{graphicx}%
\usepackage{multirow}%
\usepackage{amsmath,amssymb,amsfonts}%
\usepackage{amsthm}%
\usepackage{mathrsfs}%
\usepackage[title]{appendix}%
\usepackage{xcolor}%
\usepackage{textcomp}%
\usepackage{manyfoot}%
\usepackage{booktabs}%
\usepackage{algorithm}%
\usepackage{algorithmicx}%
\usepackage{algpseudocode}%
\usepackage{listings}%
\newcommand{\sh}{\hat \sigma}

\newcommand{\rr}{\mathbf{r}}
\newcommand{\ddd}{\mathbf{d}}
\newcommand{\RR}{\mathbf{R}}
\newcommand{\pa}[1]{\left( #1 \right)}
\newcommand{\co}[1]{\left[ #1 \right]}
\newcommand{\stx}[1]{_\text{#1}}
\begin{document}

\title[Title]{Engineering and harnessing long-range interactions for atomic quantum simulators}

\author[1]{\fnm{Javier} \sur{Arg\"uello-Luengo}}\email{javier.arguello.luengo@upc.edu}

\affil[1]{\orgdiv{Departament de Física}, \orgname{Universitat Politècnica de Catalunya}, \orgaddress{\street{Campus Nord B4-B5}, \city{Barcelona}, \postcode{08034}, \country{Spain}}}


\abstract{

    Interactions between quantum particles, such as electrons, are the source of important effects, ranging from superconductivity, to the formation of molecular bonds, or the stability of elementary compounds at high-energies. In this article, we illustrate how advances in the cold-atom community to further engineer such long-range interactions have stimulated the simulation of new regimes of these fundamental many-body problems. The goal is two-fold: first, to provide a comprehensive review of the different strategies proposed and/or experimentally realized to induce long-range interactions among atoms moving in optical potentials. Second, to showcase various fields where such platforms can offer new insights, ranging from the simulation of condensed matter phenomena to the study of lattice gauge theories, and the simulation of electronic configurations in chemistry. We then discuss the challenges and opportunities of these platforms compared to other complementary approaches based on digital simulation and quantum computation.
}

\keywords{Analog quantum simulation, long-range interactions, optical lattices}

\maketitle
\section{Introduction}\label{intro}
The potential of atomic platforms to mimic the behavior of more inaccessible quantum systems builds upon the first efforts to optically manipulate atomic systems. Thanks to the development of laser technologies during the 20th century, individual atoms trapped in laser potentials became a highly controllable quantum system that could be used to understand the quantum properties of more inaccessible systems, such as electronic ones. One of the reasons is that atoms are typically four orders of magnitude larger than electrons, which makes them easier to measure with optical techniques. Another reason is that their dynamics can be controlled with optical lattice potentials, where the atomic tunneling between neighbor sites is in the accessible range of milliseconds, which is twelve orders of magnitude slower than typical electronic effects in the attosecond scale. As a consequence, atoms soon became more favorable to measure and manipulate than electrons, and this facilitated the direct experimentation with toy-models previously related to condensed matter~\cite{Bloch2012,Gross2017}. This was the case of the first observation of a quantum phase transition between the superfluid and Mott insulating phase of the Bose-Hubbard model as the scattering length that rules on-site atomic repulsion was experimentally tuned~\cite{Greiner2002}.

Over the next 20 years, the field has evolved significantly, and the toolbox available to control itinerant atoms in optical lattices has been enriched. This includes new techniques to measure the occupation of each individual site of the lattice~\cite{hilkerRevealing2017,grossQuantum2021,Bakr2009,shersonSingleatomresolved2010} in a spin-resolved manner with quantum gas microscopes~\cite{bollSpin2016, mazurenkoColdatom2017}. More recently, the ability to create controllable interactions among atoms that are several sites apart has allowed to simulate new regimes of many-body problems that were previously inaccessible, opening new opportunities for quantum simulation. In Sec.~\ref{sec:strategies_long_range}, we review the different strategies that have been proposed to engineer long-range interactions among cold atoms. These include dipole-dipole forces, photon-mediated, or atom-mediated interactions, and we summarize the status of those platforms that have been experimentally realized. In Sec.~\ref{sec:applications}, we review different simulated problems where having access to long-range interactions is relevant, such as in the simulation of condensed matter, high-energy physics, or chemistry-related problems, focusing on those regimes that can be accessed by itinerant atoms. In Sec.~\ref{sec:perspective}, we conclude by discussing the challenges and perspectives that the engineering of controllable long-range interactions among cold atoms offers to the field of quantum simulation. Here, we will make a special emphasis on new opportunities for the simulation of chemistry-related problems.

\section{Long-range interactions among trapped atoms}
\label{sec:strategies_long_range}
Optical lattices are one of the preferred tools to control the position of atoms in ultracold experiments. When a laser beam is retoreflected by a mirror, a standing wave forms. This results in a periodic intensity profile that shifts the frequency of the atomic energy levels. Combined with atomic cooling techniques, the strength of the resulting potential $V\stx{trap}$ can exceed the atomic recoil energy, $E\stx{rec}$, and atoms get trapped at the minima of the intensity profile, where localized Wannier functions are induced. Beyond this localized trapping, the overlap between Wannier function of neighbor sites allows the atom to tunnel between adjacent sites. When the system is cooled down to the lowest band of the lattice potential, this results in a Hubbard tight-binding model of the form,
\begin{equation}
    \label{eq:Hubbard}
    \hat{H}=-J\sum_{\langle i,j\rangle}\hat{c}_i^\dag \hat{c}_j + \sum_s U_s \sum_i \hat{n}_i \hat{n}_{i+s} \,,
\end{equation}
where $\hat{n}_i=\hat{c}_i^\dag \hat{c}_i$ is the occupation number, $J$ is the tunneling rate between nearest neighbors, and $\hat{c}_i^{(\dag)}$ is the  annihilation (creation) operator of an atom at site $i$, which can be bosonic or fermionic depending on the atomic species. Due to the optical origin of the lattice, sites are separated by half a wavelength of the incoming radiation, and different lasers can be used to tailor the geometry of interest in the resulting lattice (cubic, rectangular, hexagonal...).

A crucial aspect of these simulators is the interaction of strength $U_s$ that appears between atoms separated by $s$ lattice sites. When atoms only experience contact interactions, this interatomic force decays exponentially as $U_s/U_0\sim \exp\pa{-s^2\sqrt{V\stx{trap}/E\stx{rec}}}$~\cite{zwergerMott2003,reyUltracold2004}, which limits its suitability to simulate problems where the long-range character of the interactions is relevant. As opposed to this exponential decay, here we will refer to long-range interactions as those that decay polynomially with the distance between atoms, $U_s\propto s^{-\alpha}$, with $\alpha>0$~\cite{defenuLongrange2023}. Due to the unique effects caused by these extended forces, there is an active effort to identify strategies to engineer and control them. Here, we review some of the experimentally available strategies.

\subsection{Dipole interactions}
Atoms and molecules with an effective dipole moment (electric or magnetic) interact with each other through dipole-dipole potentials of the form~\cite{CohenTannoudji1998},
\begin{equation}
\label{eq:dipoleAngle}
    V(\rr_1,\rr_2)=\frac{V_{DD}}{r^3}\co{\hat\ddd_1 \cdot \hat\ddd_2 -3(\hat\ddd_1 \cdot \hat \rr)(\hat\ddd_2 \cdot \hat \rr)}\,,
\end{equation}

The overall potential thus scales with their separation $r=|\rr_1-\rr_2|$ as $\sim r^{-3}$, and these anisotropic interactions highly depend on the relative orientation of the atomic dipoles, aligned in direction $\ddd_i$, and the unitary vector $\hat \rr$ that connects their centers [see Fig.~\ref{fig:scheme}(a)]. Still, homogeneous interactions can be identified in some reduced geometries. For example, atoms in a plane experience a homogeneous repulsive potential $V_{DD}/r^3$ when their dipole moment is oriented orthogonally to the plane. Similarly, atoms in a one-dimensional array with dipolar moments aligned along their common axis attract each other with a potential of the form, $2V_{DD}/r^3$.

The strength of the dipolar interaction is determined by the dipole moments $\mu_i$ involved, $V_{DD}=\mu_1\mu_2/(4\pi\varepsilon_0)$, whose magnitude highly depends on the physical origin of the dipole~\cite{Menotti2008}: 

\begin{figure}[!tbp]
    \centering
    \includegraphics[width=0.9\linewidth]{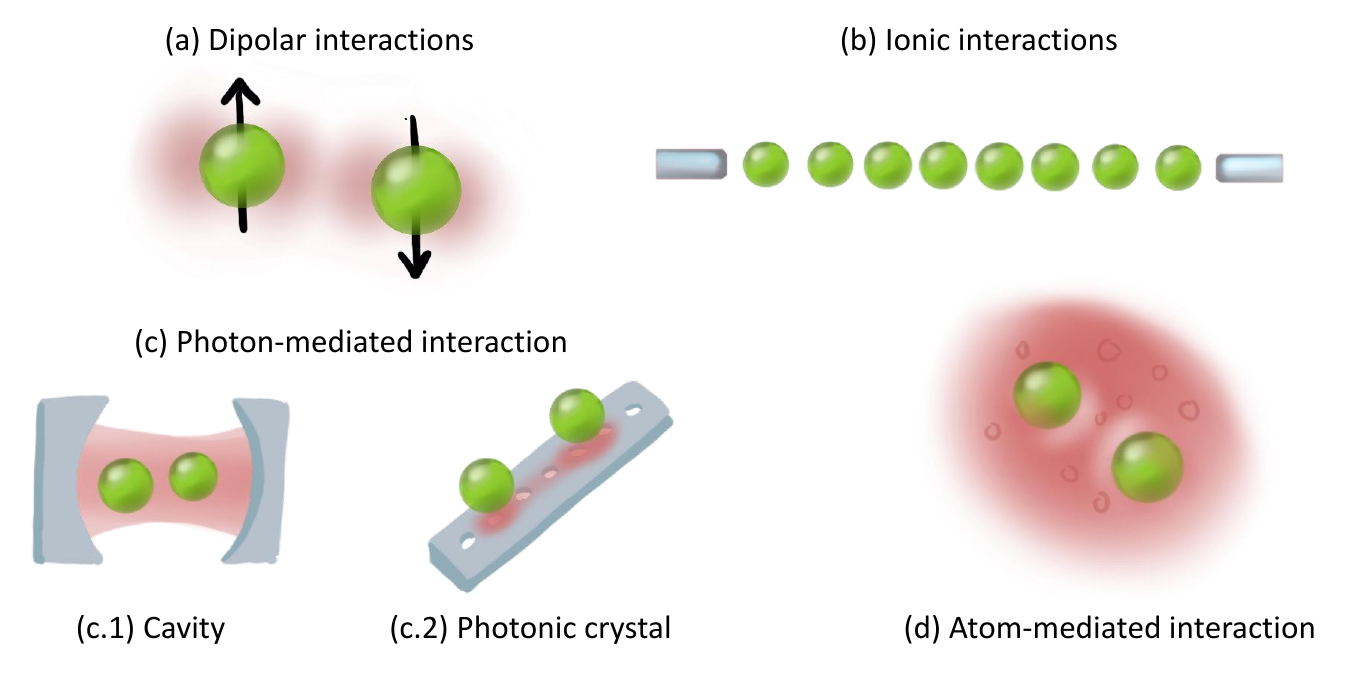}
    \caption{Schematic representation of the different strategies presented in the text to engineer long-range interactions among atomic systems. (a) Dipole-dipole interactions between polar molecules, paramagnetic atoms, or Rydberg atoms. (b) Ionic interactions between ions or ion-neutral atom mixtures. (c) Photon-mediated interactions between atoms coupled to a nanophotonic fibre (c.1), photonic crystals that engineer the dispersion relation of the mediating photons (c.2), and optical cavities that couple all atoms dressed by a cavity mode (c.3). (d) Atom-mediated interactions induced by a Fermi gas.}
    \label{fig:scheme}
    \end{figure}

\paragraph{Polar molecules}
In heteronuclear molecules, one finds net electric dipole moments in the order of a few Debyes ($\mu \sim 1D$). This leads to strong dipole interactions that can exceed typical scattering forces in cold atomic gases by two orders of magnitudes. Current techniques are capable of cooling down molecules to ultralow temperatures, which has been a longstanding challenge due to the complexity of the molecular internal structure introduced by its rotational and vibrational modes~\cite{bohn2017cold,niHigh2008,baoDipolar2023,bigagliObservation2024}. At the same time, these additional degrees of freedom, the high tunability of optical lattices and the non-linearity introduced by long-range interactions, offer promising perspectives for their use in quantum simulation and computation~\cite{Carr2009}.

\paragraph{Paramagnetic atoms}
Atomic species like Cr, or magnetic lanthanides such as Er or Dy, present large magnetic moments of order 5 $\mu_B$. While the resulting magnetic forces are typically weaker than electric dipole or contact interactions~\cite{Griesmaier2006}, the interplay with optical lattices at large filling translates into non-negligible effects, where magnetic dipole interactions compete with contact forces. For example, using $^{52}$Cr~\cite{Paz2013,Paz2016}, $^{164}$Dy~\cite{norciaTwodimensional2021}, or $^{168}$Er~\cite{aikawaBoseEinstein2012,suDipolar2023}, the role played by the spin axis and long-range interactions has been explored in many-body problems, such as the extended Bose-Hubbard model~\cite{baierExtended2016,chomazDipolar2022}.

\paragraph{Rydberg atoms}

Atoms in an excited state typically have their valence electron hundreds of \r Angstr\"oms away from their nuclei. This induces a large electric dipole that scales quadratically with the principal quantum number $n$ of this electron, which results in dipole interactions $V\stx{DD}\propto n^4$~\cite{vsibalic2018rydberg}. Additionally, a second-order dipole-dipole interaction, arising from virtual transitions to intermediate states, leads to the effective Van der Waals interaction that scales as $\sim n^{11}/r^{-6}$. For dense atomic clouds, hundreds of atoms fit within this blockade radius, and such a strong nonlinear behavior offers interesting possibilities in the construction of quantum gates~\cite{saffmanQuantum2010,Dudin2012}, memories~\cite{Distante2017,Li2016a}, or single-photon sources~\cite{Ripka2018}. The combination of these interactions with optical lattices~\cite{Anderson2011,Schauss2012} or tweezers~\cite{Browaeys2020} allows for precise control of atomic positions, enabling the simulation of relevant many-body models. In this scheme, the main source of decoherence is the reduced lifetime of the Rydberg states, on the order of 100 µs. A way to overcome this limitation is through Rydberg dressing techniques, which are currently being pursued experimentally~\cite{balewskiRydberg2014}, including for itinerant atoms in optical lattices~\cite{guardado-sanchezQuench2021}.

\subsection{Ionic systems}
While in this article we will focus on manipulating itinerant neutral atoms, it is worth mentioning the interesting physics of ions cooled down to temperatures below mK. As they are charged particles, strong electrostatic forces are induced by external electric fields. However, Gauss's law prevents a stable ion trap from being based on static electromagnetic fields (Earnshaw's theorem), which can been circumvented by other trapping techniques. For example Paul traps rely on time-averaged oscillating electric fields, while Penning traps use a combination of electric and magnetic fields~\cite{Neuhauser1980,Wineland1981}. Once at equilibrium, this trapping potential is balanced by the Coulomb repulsion between the ions, which defines crystal structures whose collective vibrational modes provide a quantum motional degree of freedom widely used in quantum simulation~\cite{blattQuantum2012, schneiderExperimental2012} [see Fig.~\ref{fig:scheme}(b)].
For example, the coupling between the internal state of the ions to the motional degrees of freedom of the crystal leads to an effective spin-spin interaction, $\sum_{i,j}J_{i,j}\sh_i \sh_j$, that exhibits a long-range scaling $J_{i,j}\propto |\rr_i-\rr_j|^{-\alpha}$ with a power-law that can be tuned between values $0<\alpha<3$~\cite{Porras2006,Britton2012}. In these traps, typical ionic separations in the order of $10$ nm result in nearest-neighbor interactions of a few $100$ kHz.
Such an experimental tunability allows one to study the role that long-range interactions can have in quantum phenomena ranging from quantum magnetism~\cite{Kim2009} to quantum transport or thermalization~\cite{Jurcevic2014}.

Going beyond the formation of ionic crystals, optical dipole traps~\cite{Schneider2010} and optical lattices~\cite{hoenigTrapping2024} have recently been adapted to ions. In this scenario, Coulomb forces among ions at neigbour lattice sites (in the range of  $100$ GHz) are typically orders of magnitude larger than optical forces . This poses some challenges, as any stray electric field needs to be detected and compensated~\cite{Karpa2019}, and ions typically separate several sites appart~\cite{Schmied2008}.
Ionic trapping in optical lattices however benefits from the scalability and versatility of these platforms~\cite{Enderlein2012}, and allows overcoming the temperature limits imposed by micromotion in electrostatic traps~\cite{Cormick2011}. This still allows for trapping lifetimes of 3 s nowadays~\cite{lambrechtLong2017}, comparable to atoms under similar conditions. 
Regarding digital quantum simulation, the motional control of these ions constitutes an interesting asset, where their extended interactions enable the implementation of quantum gates among states encoded in the ionic internal levels~\cite{Cirac2000,Lanyon2011}. By choosing a common lattice spacing, hybrid ion-atom systems are also being pushed forwards~\cite{tomzaCold2019}.

\subsection{Photon-mediated interactions}
One of the first lessons in electrodynamics is that a virtual photon exchanged between two charged particles can induce a long-range interaction between them. The more separated these particles are, the more unlikely it is for this mediating photon to propagate from one to the other, which translates into Coulomb forces that decay with the distance and whose scaling highly depends on the media the mediator propagates through.

Going beyond this virtual exchange, state-dependent interactions can be engineered through the emission and absorption of light. The challenge for this purpose is that photons emitted by an excited state rarely interact with another atom before decaying into free-space. However, by controlling the environment atoms are coupled to, one can manipulate this process, enhancing the probability that an emitted photon reaches other atoms and gaining a rich tunability of the resulting interactions~\cite{Chang2018}. In \textbf{optical cavities} with a high cooperativity, the numerous round-trips photons experience can result in an infinite-range interaction ($\alpha=0$) that effectively couples all atoms dressed by the cavity mode [see Fig.~\ref{fig:scheme}(c.1)]. The combination with optical lattices also allows one to explore the competition between short-range interactions caused by on-site collisions and the long-range interaction mediated by the cavity mode in itinerant atoms, where collective effects can stabilize self-organized supersolid phases~\cite{Baumann2010}.
The interplay of this infinite-range interaction with a magnetic field gradient aligned along the cavity axis has allowed to further tune the distance-dependence of interactions in motionally fixed atomic arrays when the intensity of the drive field is temporally modulated~\cite{Periwal2021}, which opens a new avenue for quantum simulation.

More exotic structures have also been used to mediate atom-atom interactions~\cite{Chang2018}. One example are \textbf{nanophotonic fibres}, whose radius (typically $r\sim 200$ nm) can be in the regime , $k_0r\lesssim 1$, where $k_0$ is the wave vector of the transmitted mode, which can be resonant with an atomic transition of energy $ck_0$. To satisfy the diffraction limit, this mode presents an evanescent field that extends beyond the surface of the fibre, offering a powerful resource to trap atoms and induce effective interactions mediated by a propagating photon. For example, an atom trapped in the evanescent field can directly emit a photon into the fibre mode~\cite{Goban2012a,Hinney2021}. When the photon reaches anoother trapped atom, this emission allows for an effective infinite-range spin-spin interaction that is dictated by the atomic positions, $x_i$ and $x_j$, through the phase accumulated by the mediating photon, $\propto \exp(i k_0 |x_i-x_j|)\sh_j^\dagger \sh_i$~\cite{Chang2007}. 

One of the challenges of this approach is that coherent spin exchange cannot be decoupled from collective dissipative processes. 
Applications in quantum information or quantum simulation would then highly benefit from the ability to further engineer the dispersion relation of these fibres, which is one of the main opportunities offered by photonic crystals once a periodic distribution of defects is tailored on the dielectric [see Fig.~\ref{fig:scheme}(c.2)].
Similarly to electrons propagating in periodic materials, the resulting dispersion relation shows a rich band structure, with band-gap frequencies at which photons cannot propagate~\cite{Goban2014a}. 
When the emission frequency is close enough to the band-edge, a significant propagation between nearby atoms is still possible before these exponential tails appear \cite{Douglas2015a,Gonzalez-Tudela2015}, which allows for an effective spin-spin exchange interaction over long distances $L$ among pinned atoms where fundamental many-body problems can be addressed~\cite{Hood2016a}.

\subsection{Atom-mediated interactions}
In additions to interactions mediated by a photon, a similar mediating role can be played by other type of particle or excitation. In a cold-atom experiment, a natural choice would be to rely on an additional atomic species that interacts with the simulated electrons through contact interactions and that can propagate at a faster pace. As an example, the role of the propagating photon can be played by individual atoms, excitations of a Mott insulator or excitations of a Fermi gas. While the first choice provides scalings that decay faster than Coulomb~\cite{arguello-luengoEngineering2021}, the desired $1/r$ interaction appears in the second case with the help of a cavity mode that uniformly localizes the excitation among the different simulated electrons~\cite{arguello-luengoAnalogue2019}. In the latter case of considering a fermionic gas a mediator [see Fig.~\ref{fig:scheme}(d)], the resulting effective potential follows a long-range oscillatory RKKY potential~\cite{arguello-luengoTuning2022,deFermionmediatedLongrangeInteractions2014}, as it conventionally appears in condensed matter systems. We refer the reader to an updated review~\cite{paredesPerspective2024} of the different strategies to engineer atom-mediated interactions.

\subsection{Summary of the various mediating mechanisms}
 In Table~\ref{tab:longRange} we summarize the different strategies introduced in this Section~\ref{sec:strategies_long_range} to induce long-range interactions among atoms trapped in optical potentials.  
 Their scaling ranges from the all-to-all forces present in cavities, to exponentially suppressed interactions for atoms tuned at the band-gap of photonic crystals. The strength of these atom-atom interactions at a lattice separation of $500$ nm is also diverse, ranging from a few Hz for the case of weak dipole interactions appearing in paramagnetic atoms, to several GHz in the case of strong Rydberg or ionic forces.
\hspace{-1in}
\begin{table}[!htb]
  \centering
    \begin{tabular}{cp{12em}p{6em}p{13em}}
    \toprule
    \multicolumn{2}{p{14em}}{Interactions} & Scaling & Nearest-neighbor \\
    &  & with $r$ &  strength $V_1/h$ \\
    \midrule
    \multicolumn{1}{c}{\multirow{3}[2]{*}{Dipole interactions}} & Polar molecules & $\sim r^{-3}$ & $\sim 1$ kHz  \\
          & Paramagnetic atoms & $\sim r^{-3}$ & $\sim 1$ Hz \cite{Baier2016} \\
          & Rydberg atoms & $\sim r^{-3}$ & $\sim$ 0.1 GHz $(n=20)$ \hspace{1in }$\sim$ 100 GHz $(n=40)$ \cite{vsibalic2018rydberg} \\
    \midrule
    \multicolumn{1}{c}{\multirow{2}[2]{*}{Ionic interactions}} & Ion-ion & $\sim r^{-\alpha}$,  & $\sim$ 100 kHz, (for a\\
    & & \, $\alpha \in (0,3)$ &  \quad separation $\sim$ 10 nm)~\cite{Britton2012} \\
          & Ion-neutral atom & $\sim r^{-4}$ & $\sim$ 100 Hz \cite{tomzaCold2019} \\
    \midrule
    \multicolumn{1}{c}{\multirow{3}[2]{*}{Photon-mediated }} 
          & Photonic crystals & $\sim e^{-r/L}$ & $\sim 10$ MHz~\cite{Samutpraphoot2020,Chang2018} \\
          & Optical cavities & $\sim 1$ & $\sim 0.1-1$ KHz~\cite{Periwal2021} \\
    \bottomrule
    \end{tabular}%
     \caption{Summary of the long-range scalings presented in Section~\ref{sec:strategies_long_range} for different mediating mechanisms. When not indicated otherwise, the nearest-neighbor strength is estimated for an interatomic separation of $500$ nm.}
  \label{tab:longRange}
\end{table}%

\section{Quantum simulation with long-range interactions}
\label{sec:applications}
The incorporation of extended interactions has enabled the use of atomic systems as analog simulators of physical regimes that are highly influenced by the long-range character of their dominant forces. These include early efforts in condensed-matter that have lead to recent experiments where lattice-gauge theories are simulated. More recently, technical capabilities are now pushing the possibility of entering the simulation of chemistry-related problems. 

\subsection{Condensed matter}
Long-range interactions between ultra-cold atoms are the source of many exotic many-body phenomena in condensed matter. 
For example, nearest-neighbor couplings can stabilize checker-board patterns, where atoms localize with fractional filling factors~\cite{goral02a,Bandyopadhyay2019}. These also include supersolid phases~\cite{Sinha2009,Mering2010,Buchler2003} that break translational invariance due to the formation of a crystalline structure~\cite{blandTwoDimensional2022,casottiObservation2024}. Another example are roton excitations~\cite{Mottl2012,Feng2019} that originate from the competition between attractive long-range (dipolar) and repulsive contact interactions. This can lead to the formation of quantum droplets~\cite{Ferrier-Barbut2016,Chomaz2016} (even if purely contact interaction can also stabilize them~\cite{Cabrera2018}). 
Regarding dynamics, long-range interactions in Bose-Fermi mixtures can stabilize soliton trains that maintain their shape while propagation at constant velocity~\cite{Karpiuk2004,Santhanam2006,DeSalvo2019,Cheiney2018}. Intense research has also been devoted to the study of the thermalization dynamics of these systems, and the role of many-body localization in the presence of extended interactions~\cite{nandkishoreManyBody2017, serbynQuantum2021}. 

When long-range interactions are tuned in one-dimensional systems, one can engineer topologically ordered states that are characterized by the presence of localized edge states and topological quantum numbers, as it has been observed with Rydberg atoms trapped in an array of optical tweezers~\cite{deleseleucObservation2019}. When atoms are allowed to move, such states are characterized by an uneven occupation of the sites or bonds of the lattice, that can even appear in homogeneous hamiltonians when the translational symmetry of the lattice spontaneously breaks, as in the case of bond-ordered wave phases, where the interplay between long-range order and nontrivial topology results in strongly correlated effects~\cite{arguello-luengoTuning2022,gallego-lizarribarProbing2024,baldelliFrustrated2024}.

\subsection{High-energy: lattice gauge theory}
Long range interactions have also been used in the context of Lattice Gauge Theories (LGT). One of the key interests in particle physics is understanding how particles interact via gauge degrees of freedom. For this purpose, LGT describe models in restricted geometries where a matter degree of freedom is coupled to a gauge field. 
An illustrative example is the Schwinger model. Despite the apparent simplicity of this 1+1 dimensional model for quantum electrodynamics, it shares some features with quantum chromodynamics, such as confinement and chiral symmetry breaking~\cite{chandaConfinement2020, Hauke2014}. Therefore, it has been widely adopted as a benchmark model to explore LGT techniques. To eliminate the U(1) gauge degree of freedom of this model, one needs to account for the Gauss's law that connects the electric field and charge density of the theory. This results in an exotic spin-model with two-body terms and long range interactions for the matter degree of freedom~\cite{suraceInitio2023}, which motivates the desire to induce such interactions using atomic simulators, as it was first implemented with trapped ions~\cite{martinezRealtime2016, nguyenDigital2022} and, more recently, with Rydberg atoms~\cite{gonzalez-cuadraObservation2025}. 

Other methods also rely on the use of long-range interactions to implement LGTs in optical lattices. For example, the tunable geometry of nanofabricated ionic traps~\cite{deObservation2024} or Rydberg tweezer arrays~\cite{homeierRealistic2023} can encode the matter and gauge degrees of freedom on the vertices and edges of a lattice, where extended interactions enforce the gauge constraints of the theory. Other examples are Hamiltonians where the hopping of matter from one site to the next is mediated by a gauge degree of freedom~\cite{aidelsburger15a}, which can be encoded in atomic internal levels that act as a synthetic dimension~\cite{arguello-luengoSynthetic2024,halimehColdatom2023}. Long-range Rydberg interactions can also be used to integrate out the matter degree of freedom, as proposed in Ref.~\cite{suraceLattice2020}. An updated review of LGT simulation with atoms can be found in Ref.~\cite{aidelsburgerCold2022}.

\subsection{Chemistry-related Hamiltonians}
\label{sec:quantum_chemistry}
Quantum chemistry is another problem where long-range interactions between particles appear naturally. One of the potential uses of analog simulators is thus to study the behavior of electrons in atoms and molecules. In this context, mimicking the long-range interactions among electrons and nuclei is crucial to predict the geometry of the molecular bonds they form, to simulate their dynamical evolution, or their response to external fields. 
In principle, all of these regimes reduce to solving the Schrödinger equation that describes the nuclear and electronic dynamics. However, the numerical complexity of this problem soon becomes untractable, even for moderate molecules. Over the last century, various approximations have been developed to simplify the problem numerically.

One of the most popular simplifications in quantum chemistry is the Born-Oppenheimer approximation. Given that the mass of the nuclei is three orders of magnitude larger than the mass of the electrons, their dynamics is usually much slower, which allows to solve the faster electronic problem for a fixed nuclear configuration $\{\RR_\alpha\}$. The resulting electronic problem is thus described by the Schrödinger equation:
\begin{equation}
    \label{eq:Helec}
    \hat{H}_{\text{el}}= \sum_{j=1}^{N_e}\left[-\frac{1}{2}\hat{\nabla}_j^2-\sum_{\alpha=1}^{N_n}Z_\alpha \hat{V}_c(\hat{\rr}_j,\RR_\alpha)\right]+\frac{1}{2}\sum_{i\neq j=1}^{N_e} \hat{V}_c(\hat{\rr}_i,\hat{\rr}_j)
\end{equation}
where $\hat{\rr}_j$ is the position of the $j$-th electron, $Z_\alpha$ is the charge of the $\alpha$-th nucleus, and $\hat{V}_c(\hat{\rr}_i,\hat{\rr}_j)$ is the Coulomb interaction between the electrons.

In practice, the electronic problem is often numerically solved by expanding the electronic wavefunction in a basis of Slater determinants, which are antisymmetrized products of single-electron orbitals. The accuracy of the method thus depends on the size and suitability of the basis set, which is conventionally chosen as a set of Gaussian-type orbitals to simplify the calculations. As the number of electrons increases, the number of Slater determinants grows exponentially, which makes the electronic problem untractable for large molecules~\cite{bransden2003physics}. 

Classical methods have thus relied on different approximations to simplify the problem, such the density functional theory, which approximates the exchange-correlation energy as a functional of the electron density; or Hartree-Fock methods, where the many-body wavefunction is approximated as a single Slater determinant. More refined methods have also been developed to improve the accuracy of the solution, such as the configuration interaction, the coupled cluster, or the many-body perturbation theory, which are able to capture some of the correlation effects that are neglected in the Hartree-Fock method~\cite{szaboModern1996}.

Over the last decade, an alternative approach has appeared to tackle these problems by performing the computation with quantum systems. This digital quantum computation was introduced in quantum chemistry by Aspuru-Guzik et al~\cite{Aspuru-Guzik2012}, where a Jordan-Wigner or Bravyi-Kitaev transformation maps the fermionic operators into a spin problem that is more suitable for most digital architectures. Then, quantum phase estimation or a variational quantum eigensolver extract the ground state energy of the electronic Hamiltonian~\cite{Guzik2005}. Although the initial gate complexity scaling with the number of orbitals was pessimistic (polynomial, but with a high exponent), more efficient algorithms~\cite{Berry2019a} and alternative basis sets~\cite{babbush18a,babbush18b,Low2019a}, have significantly reduced the gate complexity scaling. However, the number of qubits and the error rates of current quantum computers pose a challenge to scale this solution to large molecules.

One complementary approach to this digital strategy is thus analog simulation, which directly tackles the fermionic electronic Hamiltonian by mapping it to an appropriate fermionic system. For example, by using one simulating atom to represent each electron and/or nuclei of the molecule of interest. In the following, we will focus on different proposals that have been put forward to simulate chemistry-related problem with ultracold atoms.

\subsubsection{Lowest-energy electronic states}
Following the Born-Oppenheimer approximation, some of the first efforts in the simulation of atoms and molecules focused on the study of electronic configurations for a fixed set of nuclear positions. From the simulator, one can then extract the energy of the resulting electronic state for different nuclear configurations, which can later be used to obtain the potential energy surface of the molecule, or to infer intramoleculecular forces through the Hellmann–Feynman theorem~\cite{feynmanForces1939, hellmannEinfuehrung1944}. This provides access to information relevant to the study of chemical reactions, such as molecular bond geometry, vibrational modes, and electronic excitations.

A natural choice for the simulator is then to map each electron to an individual fermionic atom. Following this approach, simulating a system with a larger number of electrons translates into including an equally larger amount of atoms in the simulation, rather than the exponential number of multielectronic configurations that would be required in a first-principles classical calculation~\cite{szaboModern1996}. The fermionic statistics and the kinetic term in Eq.~\eqref{eq:Helec} are then already encoded by nature, and the Coulomb attraction to fixed nuclear positions corresponds to an effective potential for the atoms, which can be optically induced. In two-dimensional systems, a spatially-dependent light-shift that mimics the nuclear potential can be created with a spatial-light modulator that shapes the intensity profile of an incoming laser beam that is orthogonal to the lattice. The creation of three-dimensional intensity patterns would require additional holographic techniques, which is a technology that has already been combined with optical lattices~\cite{barredoSynthetic2018}. 

These nuclear potentials can be shaped dynamically to adiabatically prepare the electronic state of interest, and the preparation of higher energy states can benefit from Bragg transfer or amplitude modulation techniques~\cite{malzFewBody2023}. Imaging techniques such as atomic gas microscopy and time-of-flight measurement can extract the simulated electronic density and energies of the different states. The main challenge that remains is thus to engineer the electronic repulsion.

In this direction, on-site contact interactions between the simulated electrons provide a first simplified version of electronic repulsion, as illustrated in Ref.~\cite{luhmannEmulating2015} for the simulation of a benzene molecule with a hexagonal optical lattice, where each lattice site corresponds to a nucleus coordinate. Moving toward longer-range interactions, molecules, paramagnetic atoms, or Rydberg atoms naturally provide extended interactions that scale as $1/r^3$ for dipolar interactions and $1/r^6$ for van der Waals forces between Rydberg levels. Such interactions still decay faster than the $1/r$ Coulomb potential, but already capture the essential physics that makes Born-Oppenheimer chemistry problems already hard to address numerically: moving fermionic particles that feel attracted to some fixed nucleus coordinates, and experience a long-range repulsion among them. 

When electrons are mapped to individual atoms moving in optical lattices, discretization effects are defined by the electronic density on the lattice~\cite{arguello-luengoEngineering2021}. The accuracy of the simulation for a fixed number of simulated electrons is then given by the size of the lattice, which is technically constrained by the beam waist and power of the lasers used to create the lattice.
Therefore, the goal of analog simulators is not to compete with the accuracy of classical methods such as DFT or quantum Monte Carlo, but rather to serve as a complementary approach to benchmark and improve such methods in the regimes where their validity is more challenged. 
These include the study of highly correlated molecular configurations, electronic dynamics or the response to external fields.  

\subsubsection{Molecular dynamics}
Advances in laser technologies have revolutionized the understanding and control of electron dynamics on its intrinsic attosecond ($10^{-18}$ s) timescale~\cite{krauszAttosecond2009, lewensteinPrinciples2009, ciappinaAttosecond2017, salieresStudy1999}. One phenomenon that appears at this scale is high-harmonic generation (HHG), where the system absorbs multiple photons from a driving laser and emits a single photon of significantly higher energy; or non-sequential double ionization (NSDI), where two electrons are simultaneously ejected from an atom by a strong laser field.
To describe such experiments, simplified theoretical models that capture the essential dynamics have often guided the experimental realization and interpretation of these complex processes. These employ simplifications such as considering a reduced number of active electrons or ignoring the interaction between the ionized electron and the parent ion during its continuum propagation~\cite{aminiSymphony2019,eberlyHighorder1989,smirnovaMultielectron2013}. However, certain experimental regimes still require a more comprehensive description, particularly those where multielectronic processes~\cite{smirnovaMultielectron2013} or Coulomb nuclear potentials play a salient role~\cite{popruzhenkoStrong2008,popruzhenkoKeldysh2014}. This need has driven an intense development of analytical and numerical methods to extend current computational capabilities.

In this direction, analog simulators for ultrafast physics offer a highly controllable quantum system whose temporal and spatial scales are more favorable to measure than those usually found in attosecond experiments. Early proposals focused on the ionization of a single electron, whose electronic wavefunction could be mapped to the density distribution of an atomic gas in a trap~\cite{Arlinghaus2010,dumWave1998,salaUltracoldatom2017}. Given that the effective mass of the simulated electron is optically controlled, the ultrafast external field can be mimicked by magnetic or optical potentials that oscillate in the favorable regime of milliseconds, which also enables a real time imaging of the simulated electronic dynamics. The resulting electronic wavefunction can be imaged by fluorescence detection and time-of-flight techniques, as experimentally realized by Senaratne et al.~\cite{senaratneQuantum2018}. Although the oscillation of this neutral cloud does not emit radiation like charged electrons, fluorescence imaging can still be used to extract the associated dipolar acceleration and reconstruct the frequencies emitted in the HHG spectrum~\cite{arguello-luengoAnalog2024}.
For processes such as non-sequential double ionization, where more electrons become relevant to understand the observed correlated emission, one can map each electron to individual atoms in an optical trap that plays the role of the nuclear potential. For many simplified models often used to describe this NSDI process, capturing the long-range influence of the first ionized electron is a challenge. Rydberg or paramagnetic atoms can then be used to mediate the long-range interactions among the electrons, which would persist during the correlated emission of multiple electrons~\cite{arguello-luengoAnalog2024}.

\subsubsection{Beyond Born-Oppenheimer}
So far, we have reviewed applications where the Born-Oppenheimer (adiabatic) approximation is valid, so that the electronic problem can be studied for fixed nuclear positions that are simulated by external fields. There are however scenarios where the Born-Oppenheimer approximation does not hold, as it is the case of nuclear configurations where different potential energy surfaces are degenerate. Describing the molecular dynamics around these exceptional points is a challenge to most conventional numerical methods, and this offers an interesting opportunity to the benchmark offered by analog simulators. Some particular examples and strategies are the following: 
    \paragraph{Vibronics: ionic chains} 
One of the scenarios highly affected by nonadiabatic terms is photochemistry, where conical intersections enable ultrafast transitions between electronic states driven by nuclear motion, which in turn becomes strongly coupled to the electronic degrees of freedom~\cite{baer2006beyond}. This requires to account for an entangled combination of electronic configurations and nuclear vibronic states that, in the worst-case scenario, scales exponentially with the system size for conventional multiconfigurational time-dependent Hartree methods~\cite{meyerMulticonfigurational1990}. In order to simulate this molecular process, one can follow a digital encoding, where electronic states are mapped into different energy levels of ions trapped in a chain, whose vibrations provide the phononic mode where nuclear vibrations are encoded. Alternatively, this bosonic mode can also be described by photons of microwave cavities coupled to superconducting qubits in a circuit quantum electrodynamics architecture (cQED). Following this digital approach, the temporal evolution needs to be Trotterized and, interestingly, typical molecular vibronic frequencies in the 10-100 THz regime, are then mapped to the range of kHz for trapped ions, and GHz for cQED, which are more accessible to measure~\cite{MacDonell2021,olaya-agudeloSimulating2024}. As a proof of concept, photoinduced nonadiabatic dynamics has already been experimentally simulated by the team of I. Kassal~\cite{navickasExperimental2024}, where two molecular electronic states are mapped to two hyperfine levels of a single $^{171}$Yb$^{+}$ ion, and vibrations in two modes of a quadroupole trap represent two molecular vibrational modes. 

    \paragraph{Scattering experiments} Other fundamental experiments in molecular dynamics focus on the study of molecular collisions, where the trajectory of an incoming molecule or atom deviates due to the interaction with a target species. Experimentally, measuring the cross-section of these scattering trajectories provides information about the interactions that are at play. To simulate these processes, one proposal is to map nuclei and electrons to two different molecular species trapped in an optical lattice. In the presence of an external electric field, these two different molecules (or states) acquire an induced dipolar moment that can be of opposite sign. Even if the effective interaction of this toy model scales as $1/r^3$, it thus induces a repulsive force between molecules of the same kind (playing the role of electron-electron and nucleus-nucleus interactions, whose molecular dipolar moments are aligned) and an attraction between different molecules (electron-nucleus, whose molecular dipoles are antialigned). By appropriately engineering moving wavepackets that simulate the incoming projectile, this strategy allows one to investigate the cross-section of simulated scattering events beyond the Born-Oppenheimer approximation~\cite{arguello-luengoOptical2025}. One of the advantages of this approach is that atomic imaging can temporally resolve individual scattering events, unlike real experiments where those on-site measurements are not possible. This is relevant for configurations where a conical intersection is present, as the resulting cross-section is influenced by the interference of different reaction paths and how the dynamical nuclear configurations encircles the exceptional point~\cite{juanes-marcosGeometric2005,liObservation2024,kuppermannGeometric1993}. 

\section{Conclusions and perspective}
\label{sec:perspective}
In this article, we have presented different strategies that are experimentally in place to induce long-range interactions among itinerant atoms. These include the use of dipolar interactions, ions in optical traps, or the induction of photon-mediated or atom-mediated interactions, as presented in Sec.~\ref{sec:strategies_long_range}. In Sec.~\ref{sec:applications}, we have reviewed different fields where these long-range interactions are relevant for quantum simulation, including condensed matter, high-energy physics, and chemistry-related problems. 

In particular, we have focused on analog strategies where each quantum element of interest (electron, nucleus, gauge...) is mapped to an individual atom, so that the accuracy of the simulation depends on the quality and size of the system, together with the experimental control over its interactions. 
These analog simulators thus offer a complementary approach to classical methods and digital quantum computers, where the accuracy of the solution is rather determined by the number and quality of qubits and entangling gates that are accessible in a given quantum hardware. While the size and control of these digital quantum computers keeps improving, the interplay between analog and digital strategies is a promising avenue that can provide new insights on the exploration of quantum systems with extended interactions.

\section*{Acknowledgements}
We acknowledge A. González-Tudela for his insightful comments during the writing of this article, and J.I. Cirac for numerous discussions related to this work. JAL acknowledges support by the Spanish Ministerio de Ciencia e Innovaci\'on (MCIN/AEI/10.13039/501100011033, Grant No. PID2023-147469NB-C21), and by the Generalitat de Catalunya (Grant No. 2021 SGR 01411). 

\section*{Data Availability Statement}
This article does not have associated data.


\begin{thebibliography}{100}
\providecommand{\url}[1]{{#1}}
\providecommand{\urlprefix}{URL }
\providecommand{\doi}[1]{\url{https://doi.org/#1}}
\bibcommenthead

\bibitem{Bloch2012}
I.~Bloch, J.~Dalibard, S.~Nascimb{\`e}ne, Quantum simulations with ultracold
  quantum gases.
\newblock Nature Physics \textbf{8}(4), 267--276 (2012).
\newblock \doi{10.1038/nphys2259}

\bibitem{Gross2017}
C.~Gross, I.~Bloch, Quantum simulations with ultracold atoms in optical
  lattices.
\newblock Science (New York, N.Y.) \textbf{357}(6355), 995--1001 (2017).
\newblock \doi{10.1126/science.aal3837}.
\newblock {\href{https://arxiv.org/abs/28883070}{{28883070}}}

\bibitem{Greiner2002}
M.~Greiner, O.~Mandel, T.~Esslinger, T.W. H{\"a}nsch, I.~Bloch, Quantum phase
  transition from a superfluid to a {{Mott}} insulator in a gas of ultracold
  atoms.
\newblock Nature \textbf{415}(6867), 39--44 (2002).
\newblock \doi{10.1038/415039a}

\bibitem{hilkerRevealing2017}
T.A. Hilker, G.~Salomon, F.~Grusdt, A.~Omran, M.~Boll, E.~Demler, I.~Bloch,
  C.~Gross, Revealing hidden antiferromagnetic correlations in doped
  {{Hubbard}} chains via string correlators.
\newblock Science \textbf{357}(6350), 484--487 (2017).
\newblock \doi{10.1126/science.aam8990}

\bibitem{grossQuantum2021}
C.~Gross, W.S. Bakr, Quantum gas microscopy for single atom and spin detection.
\newblock Nat. Phys. \textbf{17}(12), 1316--1323 (2021).
\newblock \doi{10.1038/s41567-021-01370-5}

\bibitem{Bakr2009}
W.S. Bakr, J.I. Gillen, A.~Peng, S.~F{\"o}lling, M.~Greiner, A quantum gas
  microscope for detecting single atoms in a {{Hubbard-regime}} optical
  lattice.
\newblock Nature \textbf{462}(7269), 74--77 (2009).
\newblock \doi{10.1038/nature08482}

\bibitem{shersonSingleatomresolved2010}
J.F. Sherson, C.~Weitenberg, M.~Endres, M.~Cheneau, I.~Bloch, S.~Kuhr,
  Single-atom-resolved fluorescence imaging of an atomic {{Mott}} insulator.
\newblock Nature \textbf{467}(7311), 68--72 (2010).
\newblock \doi{10.1038/nature09378}

\bibitem{bollSpin2016}
M.~Boll, T.A. Hilker, G.~Salomon, A.~Omran, J.~Nespolo, L.~Pollet, I.~Bloch,
  C.~Gross, Spin- and density-resolved microscopy of antiferromagnetic
  correlations in {{Fermi-Hubbard}} chains.
\newblock Science \textbf{353}(6305), 1257--1260 (2016).
\newblock \doi{10.1126/science.aag1635}

\bibitem{mazurenkoColdatom2017}
A.~Mazurenko, C.S. Chiu, G.~Ji, M.F. Parsons, M.~{Kan{\'a}sz-Nagy}, R.~Schmidt,
  F.~Grusdt, E.~Demler, D.~Greif, M.~Greiner, A cold-atom
  {{Fermi}}--{{Hubbard}} antiferromagnet.
\newblock Nature \textbf{545}(7655), 462--466 (2017).
\newblock \doi{10.1038/nature22362}

\bibitem{zwergerMott2003}
W.~Zwerger, Mott--{{Hubbard}} transition of cold atoms in optical lattices.
\newblock J. Opt. B: Quantum Semiclass. Opt. \textbf{5}(2), S9 (2003).
\newblock \doi{10.1088/1464-4266/5/2/352}

\bibitem{reyUltracold2004}
A.M. Rey, Ultracold bosonic atoms in optical lattices.
\newblock Ph.D. thesis, University of Maryland (2004)

\bibitem{defenuLongrange2023}
N.~Defenu, T.~Donner, T.~Macr{\`i}, G.~Pagano, S.~Ruffo, A.~Trombettoni,
  Long-range interacting quantum systems.
\newblock Rev. Mod. Phys. \textbf{95}(3), 035002 (2023).
\newblock \doi{10.1103/RevModPhys.95.035002}

\bibitem{CohenTannoudji1998}
C.~Cohen-Tannoudji, J.~Dupont-Roc, G.~Grynberg, \emph{Atom---Photon
  Interactions} (WILEY-VCH Verlag, 1998).
\newblock \doi{10.1002/9783527617197}

\bibitem{Menotti2008}
C.~Menotti, M.~Lewenstein, T.~Lahaye, T.~Pfau, Dipolar interaction in
  ultra-cold atomic gases.
\newblock AIP Conf. Proc. \textbf{970}(1), 332 (2008).
\newblock \doi{10.1063/1.2839130}

\bibitem{bohn2017cold}
J.L. Bohn, A.M. Rey, J.~Ye, Cold molecules: {{Progress}} in quantum engineering
  of chemistry and quantum matter.
\newblock Science (80-. ). \textbf{357}(6355), 1002--1010 (2017)

\bibitem{niHigh2008}
K.K. Ni, S.~Ospelkaus, M.H.G. {de Miranda}, A.~Pe'er, B.~Neyenhuis, J.J.
  Zirbel, S.~Kotochigova, P.S. Julienne, D.S. Jin, J.~Ye, A {{High
  Phase-Space-Density Gas}} of {{Polar Molecules}}.
\newblock Science \textbf{322}(5899), 231--235 (2008).
\newblock \doi{10.1126/science.1163861}

\bibitem{baoDipolar2023}
Y.~Bao, S.S. Yu, L.~Anderegg, E.~Chae, W.~Ketterle, K.K. Ni, J.M. Doyle,
  Dipolar spin-exchange and entanglement between molecules in an optical
  tweezer array.
\newblock Science \textbf{382}(6675), 1138--1143 (2023).
\newblock \doi{10.1126/science.adf8999}

\bibitem{bigagliObservation2024}
N.~Bigagli, W.~Yuan, S.~Zhang, B.~Bulatovic, T.~Karman, I.~Stevenson, S.~Will,
  Observation of {{Bose}}--{{Einstein}} condensation of dipolar molecules.
\newblock Nature pp. 1--5 (2024).
\newblock \doi{10.1038/s41586-024-07492-z}

\bibitem{Carr2009}
L.D. Carr, D.~DeMille, R.V. Krems, J.~Ye, Cold and ultracold molecules:
  Science, technology and applications.
\newblock New J. Phys. \textbf{11}(5), 055049 (2009).
\newblock \doi{10.1088/1367-2630/11/5/055049}

\bibitem{Griesmaier2006}
A.~Griesmaier, J.~Stuhler, T.~Koch, M.~Fattori, T.~Pfau, S.~Giovanazzi,
  Comparing contact and dipolar interactions in a bose-einstein condensate.
\newblock Phys. Rev. Lett. \textbf{97}(25), 250402 (2006).
\newblock \doi{10.1103/PHYSREVLETT.97.250402/FIGURES/1/MEDIUM}

\bibitem{Paz2013}
A.~{de Paz}, A.~Sharma, A.~Chotia, E.~Mar{\'e}chal, J.H. Huckans, P.~Pedri,
  L.~Santos, O.~Gorceix, L.~Vernac, B.~{Laburthe-Tolra}, Nonequilibrium quantum
  magnetism in a dipolar lattice gas.
\newblock Phys. Rev. Lett. \textbf{111}(18), 185305 (2013).
\newblock \doi{10.1103/PhysRevLett.111.185305}

\bibitem{Paz2016}
A.~{de Paz}, P.~Pedri, A.~Sharma, M.~Efremov, B.~Naylor, O.~Gorceix,
  E.~Mar{\'e}chal, L.~Vernac, B.~{Laburthe-Tolra}, Probing spin dynamics from
  the {{Mott}} insulating to the superfluid regime in a dipolar lattice gas.
\newblock Phys. Rev. A \textbf{93}(2), 021603 (2016).
\newblock \doi{10.1103/PhysRevA.93.021603}

\bibitem{norciaTwodimensional2021}
M.A. Norcia, C.~Politi, L.~Klaus, E.~Poli, M.~Sohmen, M.J. Mark, R.N. Bisset,
  L.~Santos, F.~Ferlaino, Two-dimensional supersolidity in a dipolar quantum
  gas.
\newblock Nature \textbf{596}(7872), 357--361 (2021).
\newblock \doi{10.1038/s41586-021-03725-7}

\bibitem{aikawaBoseEinstein2012}
K.~Aikawa, A.~Frisch, M.~Mark, S.~Baier, A.~Rietzler, R.~Grimm, F.~Ferlaino,
  Bose-{{Einstein Condensation}} of {{Erbium}}.
\newblock Phys. Rev. Lett. \textbf{108}(21), 210401 (2012).
\newblock \doi{10.1103/PhysRevLett.108.210401}

\bibitem{suDipolar2023}
L.~Su, A.~Douglas, M.~Szurek, R.~Groth, S.F. Ozturk, A.~Krahn, A.H. H{\'e}bert,
  G.A. Phelps, S.~Ebadi, S.~Dickerson, F.~Ferlaino, O.~Markovi{\'c},
  M.~Greiner, Dipolar quantum solids emerging in a {{Hubbard}} quantum
  simulator.
\newblock Nature \textbf{622}(7984), 724--729 (2023).
\newblock \doi{10.1038/s41586-023-06614-3}

\bibitem{baierExtended2016}
S.~Baier, M.J. Mark, D.~Petter, K.~Aikawa, L.~Chomaz, Z.~Cai, M.~Baranov,
  P.~Zoller, F.~Ferlaino, Extended {{Bose-Hubbard}} models with ultracold
  magnetic atoms.
\newblock Science \textbf{352}(6282), 201--205 (2016).
\newblock \doi{10.1126/science.aac9812}

\bibitem{chomazDipolar2022}
L.~Chomaz, I.~{Ferrier-Barbut}, F.~Ferlaino, B.~{Laburthe-Tolra}, B.L. Lev,
  T.~Pfau, Dipolar physics: A review of experiments with magnetic quantum
  gases.
\newblock Rep. Prog. Phys. \textbf{86}(2), 026401 (2022).
\newblock \doi{10.1088/1361-6633/aca814}

\bibitem{vsibalic2018rydberg}
N.~{\v S}ibali{\'c}, C.S. Adams, \emph{Rydberg Physics} (IOP Publishing, 2018)

\bibitem{saffmanQuantum2010}
M.~Saffman, T.G. Walker, K.~M{\o}lmer, Quantum information with {{Rydberg}}
  atoms.
\newblock Rev. Mod. Phys. \textbf{82}(3), 2313--2363 (2010).
\newblock \doi{10.1103/RevModPhys.82.2313}

\bibitem{Dudin2012}
Y.O. Dudin, A.~Kuzmich, Strongly interacting {{Rydberg}} excitations of a cold
  atomic gas.
\newblock Science (80-. ). \textbf{336}(6083), 887--889 (2012).
\newblock \doi{10.1126/SCIENCE.1217901}

\bibitem{Distante2017}
E.~Distante, P.~Farrera, A.~{Padr{\'o}n-Brito}, D.~{Paredes-Barato}, G.~Heinze,
  H.~{de Riedmatten}, Storing single photons emitted by a quantum memory on a
  highly excited {{Rydberg}} state.
\newblock Nat. Comm. \textbf{8}(1), 1--6 (2017).
\newblock \doi{10.1038/ncomms14072}

\bibitem{Li2016a}
L.~Li, A.~Kuzmich, Quantum memory with strong and controllable
  {{Rydberg-level}} interactions.
\newblock Nat. Commun. \textbf{7}(1), 1--5 (2016).
\newblock \doi{10.1038/ncomms13618}

\bibitem{Ripka2018}
F.~Ripka, H.~K{\"u}bler, R.~L{\"o}w, T.~Pfau, A room-temperature single-photon
  source based on strongly interacting {{Rydberg}} atoms.
\newblock Science (80-. ). \textbf{362}(6413), 446--449 (2018).
\newblock \doi{10.1126/SCIENCE.AAU1949}

\bibitem{Anderson2011}
S.E. Anderson, K.C. Younge, G.~Raithel, Trapping rydberg atoms in an optical
  lattice.
\newblock Phys. Rev. Lett. \textbf{107}(26), 263001 (2011).
\newblock \doi{10.1103/PhysRevLett.107.263001}

\bibitem{Schauss2012}
P.~Schau{\ss}, M.~Cheneau, M.~Endres, T.~Fukuhara, S.~Hild, A.~Omran, T.~Pohl,
  C.~Gross, S.~Kuhr, I.~Bloch, Observation of spatially ordered structures in a
  two-dimensional {{Rydberg}} gas.
\newblock Nature \textbf{491}(7422), 87--91 (2012).
\newblock \doi{10.1038/nature11596}

\bibitem{Browaeys2020}
A.~Browaeys, T.~Lahaye, Many-body physics with individually controlled
  {{Rydberg}} atoms.
\newblock Nat. Phys. \textbf{16}(2), 132--142 (2020).
\newblock \doi{10.1038/s41567-019-0733-z}

\bibitem{balewskiRydberg2014}
J.B. Balewski, A.T. Krupp, A.~Gaj, S.~Hofferberth, R.~L{\"o}w, T.~Pfau, Rydberg
  dressing: Understanding of collective many-body effects and implications for
  experiments.
\newblock New J. Phys. \textbf{16}(6), 063012 (2014).
\newblock \doi{10.1088/1367-2630/16/6/063012}

\bibitem{guardado-sanchezQuench2021}
E.~{Guardado-Sanchez}, B.M. Spar, P.~Schauss, R.~Belyansky, J.T. Young,
  P.~Bienias, A.V. Gorshkov, T.~Iadecola, W.S. Bakr, Quench {{Dynamics}} of a
  {{Fermi Gas}} with {{Strong Nonlocal Interactions}}.
\newblock Phys. Rev. X \textbf{11}(2), 021036 (2021).
\newblock \doi{10.1103/PhysRevX.11.021036}

\bibitem{Neuhauser1980}
W.~Neuhauser, M.~Hohenstatt, P.E. Toschek, H.~Dehmelt, Localized visible
  {{Ba}}+ mono-ion oscillator.
\newblock Phys. Rev. A \textbf{22}(3), 1137 (1980).
\newblock \doi{10.1103/PhysRevA.22.1137}

\bibitem{Wineland1981}
D.J. Wineland, W.M. Itano, Spectroscopy of a single {{Mg}}+ ion.
\newblock Phys. Lett. A \textbf{82}(2), 75--78 (1981).
\newblock \doi{10.1016/0375-9601(81)90942-7}

\bibitem{blattQuantum2012}
R.~Blatt, C.F. Roos, Quantum simulations with trapped ions.
\newblock Nature Phys \textbf{8}(4), 277--284 (2012).
\newblock \doi{10.1038/nphys2252}

\bibitem{schneiderExperimental2012}
C.~Schneider, D.~Porras, T.~Schaetz, Experimental quantum simulations of
  many-body physics with trapped ions.
\newblock Rep. Prog. Phys. \textbf{75}(2), 024401 (2012).
\newblock \doi{10.1088/0034-4885/75/2/024401}

\bibitem{Porras2006}
D.~Porras, J.I. Cirac, Quantum manipulation of trapped ions in two dimensional
  coulomb crystals.
\newblock Phys. Rev. Lett. \textbf{96}(25), 250501 (2006).
\newblock \doi{10.1103/PhysRevLett.96.250501}

\bibitem{Britton2012}
J.W. Britton, B.C. Sawyer, A.C. Keith, C.C.J. Wang, J.K. Freericks, H.~Uys,
  M.J. Biercuk, J.J. Bollinger, Engineered two-dimensional {{Ising}}
  interactions in a trapped-ion quantum simulator with hundreds of spins.
\newblock Nature \textbf{484}(7395), 489--492 (2012).
\newblock \doi{10.1038/nature10981}

\bibitem{Kim2009}
K.~Kim, M.S. Chang, R.~Islam, S.~Korenblit, L.M. Duan, C.~Monroe, Entanglement
  and tunable spin-spin couplings between trapped ions using multiple
  transverse modes.
\newblock Phys. Rev. Lett. \textbf{103}(12), 120502 (2009).
\newblock \doi{10.1103/PhysRevLett.103.120502}

\bibitem{Jurcevic2014}
P.~Jurcevic, B.P. Lanyon, P.~Hauke, C.~Hempel, P.~Zoller, R.~Blatt, C.F. Roos,
  Quasiparticle engineering and entanglement propagation in a quantum many-body
  system.
\newblock Nature \textbf{511}(7508), 202--205 (2014).
\newblock \doi{10.1038/nature13461}

\bibitem{Schneider2010}
C.~Schneider, M.~Enderlein, T.~Huber, T.~Schaetz, Optical trapping of an ion.
\newblock Nat. Photonics \textbf{4}(11), 772--775 (2010).
\newblock \doi{10.1038/nphoton.2010.236}.
\newblock {\href{https://arxiv.org/abs/1001.2953}{{arXiv:1001.2953}}}

\bibitem{hoenigTrapping2024}
D.~Hoenig, F.~Thielemann, L.~Karpa, T.~Walker, A.~Mohammadi, T.~Schaetz,
  Trapping {{Ion Coulomb Crystals}} in an {{Optical Lattice}}.
\newblock Phys. Rev. Lett. \textbf{132}(13), 133003 (2024).
\newblock \doi{10.1103/PhysRevLett.132.133003}

\bibitem{Karpa2019}
L.~Karpa, \emph{Trapping Single Ions and Coulomb Crystals with Light Fields}
  (Springer, 2019).
\newblock \doi{10.1007/978-3-030-27716-1}

\bibitem{Schmied2008}
R.~Schmied, T.~Roscilde, V.~Murg, D.~Porras, J.I. Cirac, Quantum phases of
  trapped ions in an optical lattice.
\newblock New J. Phys. \textbf{10}(4), 045017 (2008).
\newblock \doi{10.1088/1367-2630/10/4/045017}

\bibitem{Enderlein2012}
M.~Enderlein, T.~Huber, C.~Schneider, T.~Schaetz, Single ions trapped in a
  one-dimensional optical lattice.
\newblock Phys. Rev. Lett. \textbf{109}(23), 233004 (2012).
\newblock \doi{10.1103/PhysRevLett.109.233004}.
\newblock {\href{https://arxiv.org/abs/1208.3329v1}{{arXiv:1208.3329v1}}}

\bibitem{Cormick2011}
C.~Cormick, T.~Schaetz, G.~Morigi, Trapping ions with lasers.
\newblock New J. Phys. \textbf{13}(4), 043019 (2011).
\newblock \doi{10.1088/1367-2630/13/4/043019}

\bibitem{lambrechtLong2017}
A.~Lambrecht, J.~Schmidt, P.~Weckesser, M.~Debatin, L.~Karpa, T.~Schaetz, Long
  lifetimes and effective isolation of ions in optical and electrostatic traps.
\newblock Nature Photon \textbf{11}(11), 704--707 (2017).
\newblock \doi{10.1038/s41566-017-0030-2}

\bibitem{Cirac2000}
J.I. Cirac, P.~Zoller, A scalable quantum computer with ions in an array of
  microtraps.
\newblock Nature \textbf{404}(6778), 579--581 (2000).
\newblock \doi{10.1038/35007021}

\bibitem{Lanyon2011}
B.P. Lanyon, C.~Hempel, D.~Nigg, M.~M{\"u}ller, R.~Gerritsma, F.~Z{\"a}hringer,
  P.~Schindler, J.T. Barreiro, M.~Rambach, G.~Kirchmair, M.~Hennrich,
  P.~Zoller, R.~Blatt, C.F. Roos, Universal digital quantum simulation with
  trapped ions.
\newblock Science (80-. ). \textbf{334}(6052), 57--61 (2011).
\newblock \doi{10.1126/SCIENCE.1208001/SUPPL_FILE/LANYON.SOM.PDF}.
\newblock {\href{https://arxiv.org/abs/1109.1512}{{arXiv:1109.1512}}}

\bibitem{tomzaCold2019}
M.~Tomza, K.~Jachymski, R.~Gerritsma, A.~Negretti, T.~Calarco, Z.~Idziaszek,
  P.S. Julienne, Cold hybrid ion-atom systems.
\newblock Rev. Mod. Phys. \textbf{91}(3), 035001 (2019).
\newblock \doi{10.1103/RevModPhys.91.035001}

\bibitem{Chang2018}
D.E. Chang, J.S. Douglas, A.~{Gonz{\'a}lez-Tudela}, C.L. Hung, H.J. Kimble,
  Colloquium: {{Quantum}} matter built from nanoscopic lattices of atoms and
  photons.
\newblock Reviews of Modern Physics \textbf{90}(3), 031002 (2018).
\newblock \doi{10.1103/RevModPhys.90.031002}

\bibitem{Baumann2010}
K.~Baumann, C.~Guerlin, F.~Brennecke, T.~Esslinger, Dicke quantum phase
  transition with a superfluid gas in an optical cavity.
\newblock Nature \textbf{464}(7293), 1301--1306 (2010).
\newblock \doi{10.1038/nature09009}

\bibitem{Periwal2021}
A.~Periwal, E.S. Cooper, P.~Kunkel, J.F. Wienand, E.J. Davis,
  M.~{Schleier-Smith}, Programmable interactions and emergent geometry in an
  array of atom clouds.
\newblock Nature \textbf{600}(7890), 630--635 (2021).
\newblock \doi{10.1038/s41586-021-04156-0}.
\newblock {\href{https://arxiv.org/abs/2106.04070}{{arXiv:2106.04070}}}

\bibitem{Goban2012a}
A.~Goban, K.S. Choi, D.J. Alton, D.~Ding, C.~Lacro{\^u}te, M.~Pototschnig,
  T.~Thiele, N.P. Stern, H.J. Kimble, Demonstration of a {{State-Insensitive}},
  {{Compensated Nanofiber Trap}}.
\newblock Physical Review Letters \textbf{109}(3), 033603 (2012).
\newblock \doi{10.1103/PhysRevLett.109.033603}

\bibitem{Hinney2021}
J.~Hinney, A.S. Prasad, S.~Mahmoodian, K.~Hammerer, A.~Rauschenbeutel,
  P.~Schneeweiss, J.~Volz, M.~Schemmer, Unraveling two-photon entanglement via
  the squeezing spectrum of light traveling through nanofiber-coupled atoms.
\newblock Phys. Rev. Lett. \textbf{127}(12), 123602 (2021).
\newblock \doi{10.1103/PHYSREVLETT.127.123602/FIGURES/4/MEDIUM}.
\newblock {\href{https://arxiv.org/abs/2010.09450}{{arXiv:2010.09450}}}

\bibitem{Chang2007}
D.E. Chang, A.S. S{\o}rensen, E.A. Demler, M.D. Lukin, A single-photon
  transistor using nano-scale surface plasmons.
\newblock Nat. Phys. \textbf{3}(11), 807--812 (2007).
\newblock \doi{10.1038/nphys708}.
\newblock {\href{https://arxiv.org/abs/0706.4335}{{arXiv:0706.4335}}}

\bibitem{Goban2014a}
A.~Goban, C.L. Hung, S.P. Yu, J.~Hood, J.~Muniz, J.~Lee, M.~Martin, A.~McClung,
  K.~Choi, D.~Chang, O.~Painter, H.~Kimble, Atom--light interactions in
  photonic crystals.
\newblock Nat. Comm. \textbf{5}(1), 1--9 (2014).
\newblock \doi{10.1038/ncomms4808}

\bibitem{Douglas2015a}
J.S. Douglas, H.~Habibian, C.L. Hung, A.V. Gorshkov, H.J. Kimble, D.E. Chang,
  Quantum many-body models with cold atoms coupled to photonic crystals.
\newblock Nature Photonics \textbf{9}(5), 326--331 (2015).
\newblock \doi{10.1038/nphoton.2015.57}

\bibitem{Gonzalez-Tudela2015}
A.~{Gonz{\'a}lez-Tudela}, C.L. Hung, D.E. Chang, J.I. Cirac, H.J. Kimble,
  Subwavelength vacuum lattices and atom-atom interactions in two-dimensional
  photonic crystals.
\newblock Nature Photonics \textbf{9}(5), 320--325 (2015).
\newblock \doi{10.1038/nphoton.2015.54}.
\newblock {\href{https://arxiv.org/abs/1407.7336}{{arXiv:1407.7336}}}

\bibitem{Hood2016a}
J.D. Hood, A.~Goban, A.~{Asenjo-Garcia}, M.~Lu, S.P. Yu, D.E. Chang, H.J.
  Kimble, Atom--atom interactions around the band edge of a photonic crystal
  waveguide.
\newblock Proc. Natl. Acad. Sci. \textbf{113}(38), 10507--10512 (2016).
\newblock \doi{10.1073/PNAS.1603788113}

\bibitem{arguello-luengoEngineering2021}
J.~{Arg{\"u}ello-Luengo}, T.~Shi, A.~{Gonz{\'a}lez-Tudela}, Engineering analog
  quantum chemistry {{Hamiltonians}} using cold atoms in optical lattices.
\newblock Phys. Rev. A \textbf{103}(4), 043318 (2021).
\newblock \doi{10.1103/PhysRevA.103.043318}

\bibitem{arguello-luengoAnalogue2019}
J.~{Arg{\"u}ello-Luengo}, A.~{Gonz{\'a}lez-Tudela}, T.~Shi, P.~Zoller, J.I.
  Cirac, Analogue quantum chemistry simulation.
\newblock Nature \textbf{574}(7777), 215--218 (2019).
\newblock \doi{10.1038/s41586-019-1614-4}

\bibitem{arguello-luengoTuning2022}
J.~{Arg{\"u}ello-Luengo}, A.~{Gonz{\'a}lez-Tudela}, D.~{Gonz{\'a}lez-Cuadra},
  Tuning {{Long-Range Fermion-Mediated Interactions}} in {{Cold-Atom Quantum
  Simulators}}.
\newblock Phys. Rev. Lett. \textbf{129}(8), 083401 (2022).
\newblock \doi{10.1103/PhysRevLett.129.083401}

\bibitem{deFermionmediatedLongrangeInteractions2014}
S.~De, I.B. Spielman, Fermion-mediated long-range interactions between bosons
  stored in an optical lattice.
\newblock Appl. Phys. B \textbf{114}(4), 527--536 (2014).
\newblock \doi{10.1007/s00340-013-5556-5}

\bibitem{paredesPerspective2024}
R.~Paredes, G.~Bruun, A.~{Camacho-Guardian}.
\newblock Perspective: {{Interactions}} mediated by atoms, photons, electrons,
  and excitons (2024).
\newblock \doi{10.48550/arXiv.2406.13795}

\bibitem{Baier2016}
S.~Baier, M.J. Mark, D.~Petter, K.~Aikawa, L.~Chomaz, Z.~Cai, M.~Baranov,
  P.~Zoller, F.~Ferlaino, Extended {{Bose-Hubbard}} models with ultracold
  magnetic atoms.
\newblock Science (80-. ). \textbf{352}(6282), 201--205 (2016).
\newblock \doi{10.1126/SCIENCE.AAC9812}

\bibitem{Samutpraphoot2020}
P.~Samutpraphoot, P.L. Ocola, H.~Bernien, C.~Senko, V.~Vuleti{\'c}, M.D. Lukin,
  Strong coupling of two individually controlled atoms via a nanophotonic
  cavity.
\newblock Phys. Rev. Lett. \textbf{124}(6), 063602 (2020).
\newblock \doi{10.1103/PHYSREVLETT.124.063602/FIGURES/4/MEDIUM}.
\newblock {\href{https://arxiv.org/abs/1909.09108}{{arXiv:1909.09108}}}

\bibitem{goral02a}
K.~G{\'o}ral, L.~Santos, M.~Lewenstein, Quantum phases of dipolar bosons in
  optical lattices.
\newblock Phys. Rev. Lett. \textbf{88}(17), 170406 (2002).
\newblock \doi{10.1103/PhysRevLett.88.170406}

\bibitem{Bandyopadhyay2019}
S.~Bandyopadhyay, R.~Bai, S.~Pal, K.~Suthar, R.~Nath, D.~Angom, Quantum phases
  of canted dipolar bosons in a two-dimensional square optical lattice.
\newblock Phys. Rev. A \textbf{100}(5), 053623 (2019).
\newblock \doi{10.1103/PHYSREVA.100.053623/FIGURES/5/MEDIUM}

\bibitem{Sinha2009}
S.~Sinha, K.~Sengupta, Phases and collective modes of a hardcore {{Bose-Fermi}}
  mixture in an optical lattice.
\newblock Phys. Rev. B \textbf{79}(11), 115124 (2009).
\newblock \doi{10.1103/PHYSREVB.79.115124/FIGURES/6/MEDIUM}

\bibitem{Mering2010}
A.~Mering, M.~Fleischhauer, Fermion-mediated long-range interactions of bosons
  in the one-dimensional bose-fermi-hubbard model.
\newblock Phys. Rev. A \textbf{81}(1), 11603 (2010).
\newblock \doi{10.1103/PHYSREVA.81.011603/FIGURES/4/MEDIUM}

\bibitem{Buchler2003}
H.P. B{\"u}chler, G.~Blatter, Supersolid versus phase separation in atomic
  bose-fermi mixtures.
\newblock Phys. Rev. Lett. \textbf{91}(13), 130404 (2003).
\newblock \doi{10.1103/PHYSREVLETT.91.130404/FIGURES/3/MEDIUM}

\bibitem{blandTwoDimensional2022}
T.~Bland, E.~Poli, C.~Politi, L.~Klaus, M.A. Norcia, F.~Ferlaino, L.~Santos,
  R.N. Bisset, Two-{{Dimensional Supersolid Formation}} in {{Dipolar
  Condensates}}.
\newblock Phys. Rev. Lett. \textbf{128}(19), 195302 (2022).
\newblock \doi{10.1103/PhysRevLett.128.195302}

\bibitem{casottiObservation2024}
E.~Casotti, E.~Poli, L.~Klaus, A.~Litvinov, C.~Ulm, C.~Politi, M.J. Mark,
  T.~Bland, F.~Ferlaino, Observation of vortices in a dipolar supersolid.
\newblock Nature pp. 1--5 (2024).
\newblock \doi{10.1038/s41586-024-08149-7}

\bibitem{Mottl2012}
R.~Mottl, F.~Brennecke, K.~Baumann, R.~Landig, T.~Donner, T.~Esslinger,
  Roton-type mode softening in a quantum gas with cavity-mediated long-range
  interactions.
\newblock Science \textbf{336}(6088), 1570--1573 (2012).
\newblock \doi{10.1126/science.1220314}.
\newblock {\href{https://arxiv.org/abs/1203.1322}{{arXiv:1203.1322}}}

\bibitem{Feng2019}
C.~Feng, Y.~Chen, Tunable range interactions and multi-roton excitations for
  bosons in a bose-fermi mixture with optical lattices.
\newblock Commun. Theor. Phys. \textbf{71}(7), 869 (2019).
\newblock \doi{10.1088/0253-6102/71/7/869}

\bibitem{Ferrier-Barbut2016}
I.~{Ferrier-Barbut}, H.~Kadau, M.~Schmitt, M.~Wenzel, T.~Pfau, Observation of
  quantum droplets in a strongly dipolar bose gas.
\newblock Phys. Rev. Lett. \textbf{116}(21), 215301 (2016).
\newblock \doi{10.1103/PHYSREVLETT.116.215301/FIGURES/4/MEDIUM}.
\newblock {\href{https://arxiv.org/abs/1601.03318}{{arXiv:1601.03318}}}

\bibitem{Chomaz2016}
L.~Chomaz, S.~Baier, D.~Petter, M.J. Mark, F.~W{\"a}chtler, L.~Santos,
  F.~Ferlaino, Quantum-{{Fluctuation-driven}} crossover from a dilute
  bose-einstein condensate to a macrodroplet in a dipolar quantum fluid.
\newblock Phys. Rev. X \textbf{6}(4), 41039 (2016).
\newblock \doi{10.1103/PHYSREVX.6.041039/FIGURES/5/MEDIUM}.
\newblock {\href{https://arxiv.org/abs/1607.06613}{{arXiv:1607.06613}}}

\bibitem{Cabrera2018}
C.R. Cabrera, L.~Tanzi, J.~Sanz, B.~Naylor, P.~Thomas, P.~Cheiney, L.~Tarrue,
  Quantum liquid droplets in a mixture of bose-{{Einstein}} condensates.
\newblock Science (80-. ). \textbf{359}(6373), 301--304 (2018).
\newblock \doi{10.1126/SCIENCE.AAO5686/SUPPL_FILE/AAO5686-CABRERA-SM.PDF}.
\newblock {\href{https://arxiv.org/abs/1708.07806}{{arXiv:1708.07806}}}

\bibitem{Karpiuk2004}
T.~Karpiuk, M.~Brewczyk, S.~{Ospelkaus-Schwarzer}, K.~Bongs, M.~Gajda, K.~Rz{\c
  a}zewski, Soliton trains in {{Bose-Fermi}} mixtures.
\newblock Phys. Rev. Lett. \textbf{93}(10), 100401 (2004).
\newblock \doi{10.1103/PHYSREVLETT.93.100401/FIGURES/4/MEDIUM}

\bibitem{Santhanam2006}
J.~Santhanam, V.M. Kenkre, V.V. Konotop, Solitons of {{Bose-Fermi}} mixtures in
  a strongly elongated trap.
\newblock Phys. Rev. A \textbf{73}(1), 13612 (2006).
\newblock \doi{10.1103/PHYSREVA.73.013612/FIGURES/1/MEDIUM}

\bibitem{DeSalvo2019}
B.J. DeSalvo, K.~Patel, G.~Cai, C.~Chin, Observation of fermion-mediated
  interactions between bosonic atoms.
\newblock Nature \textbf{568}(7750), 61--64 (2019).
\newblock \doi{10.1038/s41586-019-1055-0}

\bibitem{Cheiney2018}
P.~Cheiney, C.R. Cabrera, J.~Sanz, B.~Naylor, L.~Tanzi, L.~Tarruell, Bright
  soliton to quantum droplet transition in a mixture of bose-einstein
  condensates.
\newblock Phys. Rev. Lett. \textbf{120}(13), 135301 (2018).
\newblock \doi{10.1103/PHYSREVLETT.120.135301/FIGURES/4/MEDIUM}.
\newblock {\href{https://arxiv.org/abs/1710.11079}{{arXiv:1710.11079}}}

\bibitem{nandkishoreManyBody2017}
R.M. Nandkishore, S.L. Sondhi, Many-{{Body Localization}} with {{Long-Range
  Interactions}}.
\newblock Phys. Rev. X \textbf{7}(4), 041021 (2017).
\newblock \doi{10.1103/PhysRevX.7.041021}

\bibitem{serbynQuantum2021}
M.~Serbyn, D.A. Abanin, Z.~Papi{\'c}, Quantum many-body scars and weak breaking
  of ergodicity.
\newblock Nat. Phys. \textbf{17}(6), 675--685 (2021).
\newblock \doi{10.1038/s41567-021-01230-2}

\bibitem{deleseleucObservation2019}
S.~{de L{\'e}s{\'e}leuc}, V.~Lienhard, P.~Scholl, D.~Barredo, S.~Weber,
  N.~Lang, H.P. B{\"u}chler, T.~Lahaye, A.~Browaeys, Observation of a
  symmetry-protected topological phase of interacting bosons with {{Rydberg}}
  atoms.
\newblock Science \textbf{365}(6455), 775--780 (2019).
\newblock \doi{10.1126/science.aav9105}

\bibitem{gallego-lizarribarProbing2024}
K.~{Gallego-Lizarribar}, S.~{Juli{\`a}-Farr{\'e}}, M.~Lewenstein,
  C.~Weitenberg, L.~Barbiero, J.~{Arg{\"u}ello-Luengo}.
\newblock Probing spontaneously symmetry-broken phases with spin-charge
  separation through noise correlation measurements (2024).
\newblock \doi{10.48550/arXiv.2404.08374}

\bibitem{baldelliFrustrated2024}
N.~Baldelli, C.R. Cabrera, S.~{Juli{\`a}-Farr{\'e}}, M.~Aidelsburger,
  L.~Barbiero, Frustrated {{Extended Bose-Hubbard Model}} and {{Deconfined
  Quantum Critical Points}} with {{Optical Lattices}} at the {{Antimagic
  Wavelength}}.
\newblock Phys. Rev. Lett. \textbf{132}(15), 153401 (2024).
\newblock \doi{10.1103/PhysRevLett.132.153401}

\bibitem{chandaConfinement2020}
T.~Chanda, J.~Zakrzewski, M.~Lewenstein, L.~Tagliacozzo, Confinement and
  {{Lack}} of {{Thermalization}} after {{Quenches}} in the {{Bosonic Schwinger
  Model}}.
\newblock Phys. Rev. Lett. \textbf{124}(18), 180602 (2020).
\newblock \doi{10.1103/PhysRevLett.124.180602}

\bibitem{Hauke2014}
P.~Hauke, D.~Marcos, M.~Dalmonte, P.~Zoller, Quantum simulation of a lattice
  schwinger model in a chain of trapped ions.
\newblock Phys. Rev. X \textbf{3}(4), 41018 (2014).
\newblock \doi{10.1103/PhysRevX.3.041018}.
\newblock {\href{https://arxiv.org/abs/1306.2162}{{arXiv:1306.2162}}}

\bibitem{suraceInitio2023}
F.M. Surace, P.~Fromholz, N.D. Oppong, M.~Dalmonte, M.~Aidelsburger, Ab
  {{Initio Derivation}} of {{Lattice-Gauge-Theory Dynamics}} for {{Cold Gases}}
  in {{Optical Lattices}}.
\newblock PRX Quantum \textbf{4}(2), 020330 (2023).
\newblock \doi{10.1103/PRXQuantum.4.020330}

\bibitem{martinezRealtime2016}
E.A. Martinez, C.A. Muschik, P.~Schindler, D.~Nigg, A.~Erhard, M.~Heyl,
  P.~Hauke, M.~Dalmonte, T.~Monz, P.~Zoller, R.~Blatt, Real-time dynamics of
  lattice gauge theories with a few-qubit quantum computer.
\newblock Nature \textbf{534}(7608), 516--519 (2016).
\newblock \doi{10.1038/nature18318}

\bibitem{nguyenDigital2022}
N.H. Nguyen, M.C. Tran, Y.~Zhu, A.M. Green, C.H. Alderete, Z.~Davoudi, N.M.
  Linke, Digital {{Quantum Simulation}} of the {{Schwinger Model}} and
  {{Symmetry Protection}} with {{Trapped Ions}}.
\newblock PRX Quantum \textbf{3}(2), 020324 (2022).
\newblock \doi{10.1103/PRXQuantum.3.020324}

\bibitem{gonzalez-cuadraObservation2025}
D.~{Gonz{\'a}lez-Cuadra}, M.~Hamdan, T.V. Zache, B.~Braverman, M.~Kornja{\v
  c}a, A.~Lukin, S.H. Cant{\'u}, F.~Liu, S.T. Wang, A.~Keesling, M.D. Lukin,
  P.~Zoller, A.~Bylinskii, Observation of string breaking on a (2 + 1){{D
  Rydberg}} quantum simulator.
\newblock Nature pp. 1--6 (2025).
\newblock \doi{10.1038/s41586-025-09051-6}

\bibitem{deObservation2024}
A.~De, A.~Lerose, D.~Luo, F.M. Surace, A.~Schuckert, E.R. Bennewitz, B.~Ware,
  W.~Morong, K.S. Collins, Z.~Davoudi, A.V. Gorshkov, O.~Katz, C.~Monroe.
\newblock Observation of string-breaking dynamics in a quantum simulator
  (2024).
\newblock \doi{10.48550/arXiv.2410.13815}

\bibitem{homeierRealistic2023}
L.~Homeier, A.~Bohrdt, S.~Linsel, E.~Demler, J.C. Halimeh, F.~Grusdt, Realistic
  scheme for quantum simulation of
  \$\$\{\{{\textbackslash}mathbb\{\vphantom{\}\}\}}{{Z}}\vphantom\{\}\vphantom\{\}\vphantom\{\}\_\{2\}\$\$
  lattice gauge theories with dynamical matter in (2 + 1){{D}}.
\newblock Commun Phys \textbf{6}(1), 1--10 (2023).
\newblock \doi{10.1038/s42005-023-01237-6}

\bibitem{aidelsburger15a}
M.~Aidelsburger, M.~Lohse, C.~Schweizer, M.~Atala, J.T. Barreiro,
  S.~Nascimb{\`e}ne, N.R. Cooper, I.~Bloch, N.~Goldman, Measuring the {{Chern}}
  number of {{Hofstadter}} bands with ultracold bosonic atoms.
\newblock Nat. Phys. \textbf{11}(2), 162 (2015)

\bibitem{arguello-luengoSynthetic2024}
J.~{Arg{\"u}ello-Luengo}, U.~Bhattacharya, A.~Celi, R.W. Chhajlany, T.~Grass,
  M.~P{\l}odzie{\'n}, D.~Rakshit, T.~Salamon, P.~Stornati, L.~Tarruell,
  M.~Lewenstein, Synthetic dimensions for topological and quantum phases.
\newblock Commun Phys \textbf{7}(1), 1--10 (2024).
\newblock \doi{10.1038/s42005-024-01636-3}

\bibitem{halimehColdatom2023}
J.C. Halimeh, M.~Aidelsburger, F.~Grusdt, P.~Hauke, B.~Yang.
\newblock Cold-atom quantum simulators of gauge theories (2023).
\newblock \doi{10.48550/arXiv.2310.12201}

\bibitem{suraceLattice2020}
F.M. Surace, P.P. Mazza, G.~Giudici, A.~Lerose, A.~Gambassi, M.~Dalmonte,
  Lattice {{Gauge Theories}} and {{String Dynamics}} in {{Rydberg Atom Quantum
  Simulators}}.
\newblock Phys. Rev. X \textbf{10}(2), 021041 (2020).
\newblock \doi{10.1103/PhysRevX.10.021041}

\bibitem{aidelsburgerCold2022}
M.~Aidelsburger, L.~Barbiero, A.~Bermudez, T.~Chanda, A.~Dauphin,
  D.~{Gonz{\'a}lez-Cuadra}, P.R. Grzybowski, S.~Hands, F.~Jendrzejewski,
  J.~J{\"u}nemann, G.~Juzeli{\=u}nas, V.~Kasper, A.~Piga, S.J. Ran, M.~Rizzi,
  G.~Sierra, L.~Tagliacozzo, E.~Tirrito, T.V. Zache, J.~Zakrzewski, E.~Zohar,
  M.~Lewenstein, Cold atoms meet lattice gauge theory.
\newblock Philosophical Transactions of the Royal Society A: Mathematical,
  Physical and Engineering Sciences \textbf{380}(2216), 20210064 (2022).
\newblock \doi{10.1098/rsta.2021.0064}

\bibitem{bransden2003physics}
B.H. Bransden, C.J. Joachain, T.J. Plivier, \emph{Physics of Atoms and
  Molecules} (Pearson education, 2003)

\bibitem{szaboModern1996}
A.~Szabo, N.S. Ostlund, \emph{Modern Quantum Chemistry: Introduction to
  Advanced Electronic Structure Theory} (Courier Corporation, 1996)

\bibitem{Aspuru-Guzik2012}
A.~{Aspuru-Guzik}, P.~Walther, Photonic quantum simulators.
\newblock Nature Physics \textbf{8} (2012).
\newblock \doi{10.1038/NPHYS2253}

\bibitem{Guzik2005}
A.~{Aspuru-Guzik}, A.D. Dutoi, P.J. Love, M.~{Head-Gordon}, Simulated {{Quantum
  Computation}} of {{Molecular Energies}}.
\newblock Science \textbf{309}(5741) (2005).
\newblock \doi{10.1126/science.1113479}

\bibitem{Berry2019a}
D.W. Berry, C.~Gidney, M.~Motta, J.R. McClean, R.~Babbush, Qubitization of
  arbitrary basis quantum chemistry leveraging sparsity and low rank
  factorization.
\newblock Quantum \textbf{3}, 208 (2019).
\newblock \doi{10.22331/q-2019-12-02-208}.
\newblock {\href{https://arxiv.org/abs/1902.02134}{{arXiv:1902.02134}}}

\bibitem{babbush18a}
R.~Babbush, N.~Wiebe, J.~McClean, J.~McClain, H.~Neven, G.K.L. Chan, Low-depth
  quantum simulation of materials.
\newblock Phys. Rev. X \textbf{8}(1), 11044 (2018).
\newblock \doi{10.1103/PhysRevX.8.011044}

\bibitem{babbush18b}
R.~Babbush, D.W. Berry, J.R. McClean, H.~Neven, Quantum simulation of chemistry
  with sublinear scaling to the continuum.
\newblock arXiv:1807.09802  (2018).
\newblock {\href{https://arxiv.org/abs/1807.09802}{{arXiv:1807.09802}}}

\bibitem{Low2019a}
G.H. Low, I.L. Chuang, Hamiltonian simulation by qubitization.
\newblock Quantum \textbf{3}, 163 (2019).
\newblock \doi{10.22331/q-2019-07-12-163}.
\newblock {\href{https://arxiv.org/abs/1610.06546}{{arXiv:1610.06546}}}

\bibitem{feynmanForces1939}
R.P. Feynman, Forces in {{Molecules}}.
\newblock Phys. Rev. \textbf{56}(4), 340--343 (1939).
\newblock \doi{10.1103/PhysRev.56.340}

\bibitem{hellmannEinfuehrung1944}
H.~Hellmann, \emph{{Einf{\"u}hrung in die Quantenchemie}} (J.W. Edwards, Ann
  Arbor, Mich, 1944)

\bibitem{barredoSynthetic2018}
D.~Barredo, V.~Lienhard, S.~{de L{\'e}s{\'e}leuc}, T.~Lahaye, A.~Browaeys,
  Synthetic three-dimensional atomic structures assembled atom by atom.
\newblock Nature \textbf{561}(7721), 79--82 (2018).
\newblock \doi{10.1038/s41586-018-0450-2}

\bibitem{malzFewBody2023}
D.~Malz, J.I. Cirac, Few-{{Body Analog Quantum Simulation}} with
  {{Rydberg-Dressed Atoms}} in {{Optical Lattices}}.
\newblock PRX Quantum \textbf{4}(2), 020301 (2023).
\newblock \doi{10.1103/PRXQuantum.4.020301}

\bibitem{luhmannEmulating2015}
D.S. L{\"u}hmann, C.~Weitenberg, K.~Sengstock, Emulating {{Molecular Orbitals}}
  and {{Electronic Dynamics}} with {{Ultracold Atoms}}.
\newblock Phys. Rev. X \textbf{5}(3), 031016 (2015).
\newblock \doi{10.1103/PhysRevX.5.031016}

\bibitem{krauszAttosecond2009}
F.~Krausz, M.~Ivanov, Attosecond physics.
\newblock Rev. Mod. Phys. \textbf{81}(1), 163--234 (2009).
\newblock \doi{10.1103/RevModPhys.81.163}

\bibitem{lewensteinPrinciples2009}
M.~Lewenstein, A.~L'Huillier, in \emph{Strong {{Field Laser Physics}}}, ed. by
  T.~Brabec, Springer {{Series}} in {{Optical Sciences}} (Springer, New York,
  NY, 2009), pp. 147--183.
\newblock \doi{10.1007/978-0-387-34755-4_7}

\bibitem{ciappinaAttosecond2017}
M.F. Ciappina, J.A. {P{\'e}rez-Hern{\'a}ndez}, A.S. Landsman, W.A. Okell,
  S.~Zherebtsov, B.~F{\"o}rg, J.~Sch{\"o}tz, L.~Seiffert, T.~Fennel,
  T.~Shaaran, T.~Zimmermann, A.~Chac{\'o}n, R.~Guichard, A.~Za{\"i}r, J.W.G.
  Tisch, J.P. Marangos, T.~Witting, A.~Braun, S.A. Maier, L.~Roso,
  M.~Kr{\"u}ger, P.~Hommelhoff, M.F. Kling, F.~Krausz, M.~Lewenstein,
  Attosecond physics at the nanoscale.
\newblock Rep. Prog. Phys. \textbf{80}(5), 054401 (2017).
\newblock \doi{10.1088/1361-6633/aa574e}

\bibitem{salieresStudy1999}
P.~Sali{\`e}res, A.~L'Huillier, P.~Antoine, M.~Lewenstein, in \emph{Advances
  {{In Atomic}}, {{Molecular}}, and {{Optical Physics}}}, vol.~41, ed. by
  B.~Bederson, H.~Walther (Academic Press, 1999), pp. 83--142.
\newblock \doi{10.1016/S1049-250X(08)60219-0}

\bibitem{aminiSymphony2019}
K.~Amini, J.~Biegert, F.~Calegari, A.~Chac{\'o}n, M.F. Ciappina, A.~Dauphin,
  D.K. Efimov, C.~{Figueira de Morisson Faria}, K.~Giergiel, P.~Gniewek, A.S.
  Landsman, M.~Lesiuk, M.~Mandrysz, A.S. Maxwell, R.~Moszy{\'n}ski, L.~Ortmann,
  J.~{Antonio P{\'e}rez-Hern{\'a}ndez}, A.~Pic{\'o}n, E.~Pisanty,
  J.~{Prauzner-Bechcicki}, K.~Sacha, N.~Su{\'a}rez, A.~Za{\"i}r, J.~Zakrzewski,
  M.~Lewenstein, Symphony on strong field approximation.
\newblock Rep Prog Phys \textbf{82}(11), 116001 (2019).
\newblock \doi{10.1088/1361-6633/ab2bb1}

\bibitem{eberlyHighorder1989}
J.H. Eberly, Q.~Su, J.~Javanainen, High-order harmonic production in
  multiphoton ionization.
\newblock J. Opt. Soc. Am. B, JOSAB \textbf{6}(7), 1289--1298 (1989).
\newblock \doi{10.1364/JOSAB.6.001289}

\bibitem{smirnovaMultielectron2013}
O.~Smirnova, M.~Ivanov.
\newblock Multielectron {{High Harmonic Generation}}: Simple man on a complex
  plane, chapter 7 in {{Attosecond}} and {{XUV}} physics, edited by {{T}}.
  {{Schultz}} and {{M}}. {{Vrakking}} (2013)

\bibitem{popruzhenkoStrong2008}
S.~Popruzhenko, D.~Bauer, Strong field approximation for systems with
  {{Coulomb}} interaction.
\newblock Journal of Modern Optics \textbf{55}(16), 2573--2589 (2008).
\newblock \doi{10.1080/09500340802161881}

\bibitem{popruzhenkoKeldysh2014}
S.V. Popruzhenko, Keldysh theory of strong field ionization: History,
  applications, difficulties and perspectives.
\newblock J. Phys. B: At. Mol. Opt. Phys. \textbf{47}(20), 204001 (2014).
\newblock \doi{10.1088/0953-4075/47/20/204001}

\bibitem{Arlinghaus2010}
S.~Arlinghaus, M.~Holthaus, Driven optical lattices as strong-field simulators.
\newblock Physical Review A \textbf{81}(6), 063612 (2010).
\newblock \doi{10.1103/PhysRevA.81.063612}

\bibitem{dumWave1998}
R.~Dum, A.~Sanpera, K.A. Suominen, M.~Brewczyk, M.~Ku{\'s}, K.~Rzazewski,
  M.~Lewenstein, Wave {{Packet Dynamics}} with {{Bose-Einstein Condensates}}.
\newblock Phys. Rev. Lett. \textbf{80}(18), 3899--3902 (1998).
\newblock \doi{10.1103/PhysRevLett.80.3899}

\bibitem{salaUltracoldatom2017}
S.~Sala, J.~F{\"o}rster, A.~Saenz, Ultracold-atom quantum simulator for
  attosecond science.
\newblock Phys. Rev. A \textbf{95}(1), 011403 (2017).
\newblock \doi{10.1103/PhysRevA.95.011403}

\bibitem{senaratneQuantum2018}
R.~Senaratne, S.V. Rajagopal, T.~Shimasaki, P.E. Dotti, K.M. Fujiwara,
  K.~Singh, Z.A. Geiger, D.M. Weld, Quantum simulation of ultrafast dynamics
  using trapped ultracold atoms.
\newblock Nat Commun \textbf{9}(1), 2065 (2018).
\newblock \doi{10.1038/s41467-018-04556-3}

\bibitem{arguello-luengoAnalog2024}
J.~{Arg{\"u}ello-Luengo}, J.~{Rivera-Dean}, P.~Stammer, A.S. Maxwell, D.M.
  Weld, M.F. Ciappina, M.~Lewenstein, Analog {{Simulation}} of {{High-Harmonic
  Generation}} in {{Atoms}}.
\newblock PRX Quantum \textbf{5}(1), 010328 (2024).
\newblock \doi{10.1103/PRXQuantum.5.010328}

\bibitem{baer2006beyond}
M.~Baer, \emph{Beyond {{Born-Oppenheimer}}: Electronic Nonadiabatic Coupling
  Terms and Conical Intersections} (John Wiley \& Sons, 2006)

\bibitem{meyerMulticonfigurational1990}
H.D. Meyer, U.~Manthe, L.S. Cederbaum, The multi-configurational time-dependent
  {{Hartree}} approach.
\newblock Chemical Physics Letters \textbf{165}(1), 73--78 (1990).
\newblock \doi{10.1016/0009-2614(90)87014-I}

\bibitem{MacDonell2021}
R.J. MacDonell, C.E. Dickerson, C.J. Birch, A.~Kumar, C.L. Edmunds, M.J.
  Biercuk, C.~Hempel, I.~Kassal, Analog quantum simulation of chemical
  dynamics.
\newblock Chem. Sci. \textbf{12}(28), 9794--9805 (2021).
\newblock \doi{10.1039/D1SC02142G}.
\newblock {\href{https://arxiv.org/abs/2012.01852}{{arXiv:2012.01852}}}

\bibitem{olaya-agudeloSimulating2024}
V.C. {Olaya-Agudelo}, B.~Stewart, C.H. Valahu, R.J. MacDonell, M.J. Millican,
  V.G. Matsos, F.~Scuccimarra, T.R. Tan, I.~Kassal.
\newblock Simulating open-system molecular dynamics on analog quantum computers
  (2024).
\newblock \doi{10.48550/arXiv.2407.17819}

\bibitem{navickasExperimental2024}
T.~Navickas, R.J. MacDonell, C.H. Valahu, V.C. {Olaya-Agudelo}, F.~Scuccimarra,
  M.J. Millican, V.G. Matsos, H.L. Nourse, A.D. Rao, M.J. Biercuk, C.~Hempel,
  I.~Kassal, T.R. Tan.
\newblock Experimental {{Quantum Simulation}} of {{Chemical Dynamics}} (2024).
\newblock \doi{10.48550/arXiv.2409.04044}

\bibitem{arguello-luengoOptical2025}
J.~{Arg{\"u}ello-Luengo}, A.~{Gonz{\'a}lez-Tudela}, J.I. Cirac.
\newblock Optical lattice quantum simulator of dynamics beyond
  {{Born-Oppenheimer}} (2025).
\newblock \doi{10.48550/arXiv.2503.23464}

\bibitem{juanes-marcosGeometric2005}
J.C. {Juanes-Marcos}, S.C. Althorpe, Geometric phase effects in the
  {{H}}+{{H2}} reaction: {{Quantum}} wave-packet calculations of integral and
  differential cross sections.
\newblock The Journal of Chemical Physics \textbf{122}(20), 204324 (2005).
\newblock \doi{10.1063/1.1924411}

\bibitem{liObservation2024}
S.~Li, J.~Huang, Z.~Lu, Y.~Shu, W.~Chen, D.~Yuan, T.~Wang, B.~Fu, Z.~Zhang,
  X.~Wang, D.H. Zhang, X.~Yang, Observation of geometric phase effect through
  backward angular oscillations in the {{H}} + {{HD}} {$\rightarrow$} {{H2}} +
  {{D}} reaction.
\newblock Nat Commun \textbf{15}(1), 1698 (2024).
\newblock \doi{10.1038/s41467-024-45843-6}

\bibitem{kuppermannGeometric1993}
A.~Kuppermann, Y.S.M. Wu, The geometric phase effect shows up in chemical
  reactions.
\newblock Chemical Physics Letters \textbf{205}(6), 577--586 (1993).
\newblock \doi{10.1016/0009-2614(93)80015-H}

\end{thebibliography}
\end{document}